\documentclass[letterpaper,notoc]{CECS}

\title{Exact black hole solution with a minimally coupled scalar field}
\author{Cristi\'{a}n Mart\'{\i }nez, Ricardo Troncoso and Jorge Zanelli\footnote{{\it E-mail:} {\tt  martinez@cecs.cl, ratron@cecs.cl, jz@cecs.cl}}\\ Centro de Estudios Cient\'{\i}ficos (CECS), Casilla 1469, Valdivia, Chile.  }

\preprint{{\tiny CECS-PHY-04/12} }

\abstract{
An exact four-dimensional black hole solution of gravity with a minimally
coupled self-interacting scalar field is reported. The event horizon is a
surface of negative constant curvature enclosing the curvature singularity
at the origin, and the scalar field is regular everywhere outside the
origin. This solution is an asymptotically locally AdS spacetime. The strong
energy condition is satisfied on and outside the event horizon. The
thermodynamical analysis shows the existence of a critical temperature,
below which a black hole in vacuum undergoes a spontaneous dressing up with
a nontrivial scalar field in a process reminiscent of ferromagnetism.  }

\begin{document}
\section{Introduction}

In four dimensions, exact black hole solutions in General Relativity with a
minimally coupled self-interacting scalar field have not been reported
previously. This fact may be seen as a natural consequence of the so called
no-hair conjecture, which originally stated that a black hole should be
characterized only in terms of its mass, angular momentum and electric
charge \cite{Ruffini-Wheeler,Bekenstein:1971hc} (for recent discussions see
e.g., \cite{review}). This obstacle can be circumvented introducing a
cosmological constant $\Lambda $ and a conformal coupling, in which case
exact black hole solutions are known in three \cite
{Martinez:1996gn,Henneaux:2002wm} and four dimensions \cite{Martinez:2002ru}%
. For vanishing $\Lambda $, a four-dimensional black hole is also known, but
the scalar field diverges at the horizon \cite{BBMB}.

However, for minimal coupling, an exact black hole solution is known only in
three dimensions \cite{Henneaux:2002wm}, provided $\Lambda <0$, in which
case, spherically symmetric black hole solutions have also been found
numerically in four \cite{Torii:2001pg,Winstanley:2002jt} and five
dimensions \cite{Hertog:2004dr}. On the other hand, a negative cosmological
also allows for the existence of black holes whose horizon has nontrivial
topology in four \cite{Lemos, Vanzo:1997gw,Brill:1997mf} and higher
dimensions \cite{Birmingham,Cai-Soh} as well as for gravity theories
containing higher powers of the curvature \cite
{BHscan,Aros:2000ij,Cai-GB,Dehghani}.

In this paper we report an exact black hole solution in four-dimensions for
gravity with a negative cosmological constant, with a minimally coupled
self-interacting scalar field. The event horizon is a surface of negative
constant curvature enclosing the curvature singularity at the origin. The
spacetime is asymptotically locally anti-de Sitter (AdS), and the scalar
field is regular everywhere outside the origin. It is shown that there is a
second order phase transition at a critical temperature $T_{c}=(2\pi l)^{-1}$%
, below which a black hole in vacuum undergoes a spontaneous dressing up
with a nontrivial scalar field. As is shown in the appendix, the
transformation of this theory to the conformal frame maps this solution into
another black hole with a rich causal structure.

\section{Black hole solution}

Consider four-dimensional gravity with negative cosmological constant ($%
\Lambda =-3l^{-2}$) and a scalar field described by the action 
\begin{equation}
I[g_{\mu \nu },\phi ]=\int d^{4}x\sqrt{-g}\left[ \frac{R+6l^{-2}}{16\pi G}-%
\frac{1}{2}g^{\mu \nu }\partial _{\mu }\phi \partial _{\nu }\phi -V(\phi
)\right] \;,  \label{action}
\end{equation}
where $l$ is AdS radius, and $G$ is Newton's constant. We take the following
self-interaction potential 
\begin{equation}
V(\phi )=-\frac{3}{4\pi Gl^{2}}\sinh ^{2}\sqrt{\frac{4\pi G}{3}}\phi \;,
\label{potential}
\end{equation}
which has a global maximum at $\phi =0$, and has a mass term given by $%
m^{2}=\left. V^{\prime \prime }\right| _{\phi =0}=-2l^{-2}$. This mass
satisfies the Breitenlohner-Friedman bound that ensures the perturbative
stability of AdS spacetime \cite{B-F,M-T}. This potential has a natural
interpretation in the conformal frame (see the Appendix). The field
equations are 
\begin{eqnarray}
G_{\mu \nu }-\frac{3}{l^{2}}g_{\mu \nu } &=&8\pi G\,T_{\mu \nu }\;,
\label{Eeq} \\
\square \phi -\frac{dV}{d\phi } &=&0\;,  \label{Feq}
\end{eqnarray}
where $\square \equiv g^{\mu \nu }\nabla _{\mu }\nabla _{\nu }$, and the
stress-energy tensor is given by 
\begin{equation}
T_{\mu \nu }=\partial _{\mu }\phi \partial _{\nu }\phi -\frac{1}{2}g_{\mu
\nu }g^{\alpha \beta }\partial _{\alpha }\phi \partial _{\beta }\phi -g_{\mu
\nu }V(\phi )\;.  \label{Tuv}
\end{equation}

A static black hole solution with topology $\Bbb{R}^{2}\times \Sigma $,
where $\Sigma $ is a two-dimensional manifold of negative constant
curvature, is given by 
\begin{equation}
ds^{2}=\frac{r(r+2G\mu )}{(r+G\mu )^{2}}\left[ -\left( \frac{r^{2}}{l^{2}}%
-\left( 1+\frac{G\mu }{r}\right) ^{2}\right) dt^{2}+\left( \frac{r^{2}}{l^{2}%
}-\left( 1+\frac{G\mu }{r}\right) ^{2}\right) ^{-1}dr^{2}+r^{2}d\sigma
^{2}\right] \;,  \label{Black-Hole}
\end{equation}
and the scalar field is 
\begin{equation}
\phi =\sqrt{\frac{3}{4\pi G}}\;\mbox{Arctanh}\frac{G\mu }{r+G\mu }\;.
\label{scalar}
\end{equation}
Here $d\sigma ^{2}$ is the line element of the base manifold $\Sigma $,
which has negative constant curvature (rescaled to $-1$) and hence is
locally isometric to the hyperbolic manifold $H^{2}$. This means that $%
\Sigma $ must be of the form $\Sigma =H^{2}/\Gamma $, where $\Gamma $ is a
freely acting discrete subgroup of $O(2,1)$ (i.e., without fixed points).
The configurations (\ref{Black-Hole}) are asymptotically locally AdS
spacetimes, with a single timelike Killing vector $\partial _{t}$, provided $%
\Sigma $ is assumed to be compact without boundary\footnote{%
For $\mu =0$, these spacetimes admit Killing spinors provided $\Sigma $ is a
noncompact surface \cite{Aros:2002rk}. Such a configuration describes the
supersymmetric ground state of a warped black string and is therefore
expected to be stable. As it was shown in Ref. \cite{Gibbons:2002pq}, these
configurations are also stable under gravitational perturbations.}. As it is
shown in section \ref{thermo}, the mass of this solution is given by 
\begin{equation}
M=\frac{\sigma }{4\pi }\mu \;,  \label{BlackHoleMass}
\end{equation}
where $\sigma $ denotes the area of $\Sigma $. The only singularities of the
curvature and the scalar field occur where the conformal factor in (\ref
{Black-Hole}) vanishes, that is at $r=0$ and at $r=-2G\mu $. The range of $r$
is taken as $r>-2G\mu $ for negative mass, and $r>0$ otherwise. These
singularities are surrounded by an event horizon located at 
\begin{equation}
r_{+}=\frac{l}{2}\left( 1+\sqrt{1+4G\mu /l}\right) \;,  \label{horizon}
\end{equation}
provided the mass is bounded from below by 
\begin{equation}
\mu >-\frac{l}{4G}\;.  \label{Mass-range}
\end{equation}
The causal structure is the same as for the Schwarzschild-AdS black hole,
but at each point of the Penrose diagram the sphere is replaced by $\Sigma $.

For non-negative masses, the horizon radius satisfies $r_{+}\geq l$. For the
massless case, the metric takes a simple form 
\begin{equation}
d\bar{s}^{2}=-\left( \frac{r^{2}}{l^{2}}-1\right) dt^{2}+\left( \frac{r^{2}}{%
l^{2}}-1\right) ^{-1}dr^{2}+r^{2}d\sigma ^{2}\,,  \label{muzero}
\end{equation}
which is a locally AdS spacetime (\emph{i.e.} it has constant curvature),
and the scalar field vanishes. For $-l(4G)^{-1}<\mu<0$, the horizon radius
is in the range $l/2<r_+<l$, and for $\mu =-l(4G)^{-1}$, $r_{+}=l/2$ becomes
a double root which coincides with the singularities.

Note that the scalar field cannot be switched off keeping the mass fixed. In
fact, there is only one integration constant ($\mu $), and for $\phi
\rightarrow 0$ the geometry approaches the massless black hole (\ref{muzero}%
). This means that for a given mass there are two branches of different
black hole solutions, the one with non-trivial scalar field (\ref{Black-Hole}%
, \ref{scalar}), and the vacuum solution (with $\phi =0$) found in Refs. 
\cite{Vanzo:1997gw, Brill:1997mf}, whose metric reads 
\begin{equation}
ds_{0}^{2}=-\left[ \frac{\rho ^{2}}{l^{2}}-1-\frac{2G\mu _{0}}{\rho }\right]
dt^{2}+\left[ \frac{\rho ^{2}}{l^{2}}-1-\frac{2G\mu _{0}}{\rho }\right]
^{-1}d\rho ^{2}+\rho ^{2}d\sigma ^{2}\;,  \label{phizero}
\end{equation}
with mass $M=\sigma \mu _{0}/4\pi $.

\section{Thermodynamics}

\label{thermo}

The partition function for a thermodynamical ensemble is identified with the
Euclidean path integral in the saddle point approximation around the
Euclidean continuation of the classical solution \cite{Gibbons:1976ue}.
Thus, we consider the Euclidean continuation of the action (\ref{action}) in
Hamiltonian form 
\begin{equation}
I=\int \left[ \pi ^{ij}\dot{g}_{ij}+p\dot{\phi}-N\mathcal{H}-N^{i}\mathcal{H}%
_{i}\right] d^{3}xdt+B,  \label{haction}
\end{equation}
where $B$ is a surface term. We consider the minisuperspace of static
Euclidean metrics 
\begin{equation}
ds^{2}=N(r)^{2}f(r)^{2}dt^{2}+f(r)^{-2}dr^{2}+R(r)^{2}d\sigma ^{2}
\label{dse}
\end{equation}
with $0\leq t\leq \beta $ periodic, $r\ge r_{+}$ and a scalar field of the
form $\phi =\phi (r)$. The inverse of the period $\beta $ corresponds to the
temperature $T$, and the reduced Hamiltonian action is 
\begin{equation}
I=-\beta \,\sigma \int_{r_{+}}^{\infty }N(r)\mathcal{H}(r)dr+B,
\label{rhaction}
\end{equation}
where $\sigma $ is the area of the base manifold $\Sigma $, and 
\begin{eqnarray*}
\mathcal{H} &=&NR^{2}\left[ \frac{1}{8\pi G}\left( \frac{(f^{2})^{\prime
}R^{\prime }}{R}+\frac{2f^{2}R^{\prime \prime }}{R}+\frac{1}{R^{2}}(1
+f^{2})\right) \right. \\
&&\left. +\Lambda +\frac{1}{2}f^{2}(\phi ^{\prime })^{2}+V(\phi )\right] .
\end{eqnarray*}

The Euclidean black hole solution is static and satisfies the constraint $%
\mathcal{H}=0$. Therefore, the value of the action on the classical solution
is just given by the boundary term $B$. This boundary term must be such that
the action (\ref{haction}) attains an extremum within the class of fields
considered here \cite{Regge:1974zd}. We now turn to the evaluation of the
Euclidean action on shell. The condition that the geometries allowed in the
variation should contain no conical singularities at the horizon requires 
\begin{equation}
\left. \beta (N(r)(f^{2}(r))^{\prime })\right| _{r=r_{+}}=4\pi \;,
\label{regular}
\end{equation}
which for the solution (\ref{Black-Hole}), directly yields the period $\beta 
$ as a function of $r_{+}$, 
\begin{equation}
T=\beta ^{-1}=\frac{1}{2\pi l}\left( \frac{2r_{+}}{l}-1\right) \;.
\label{beta}
\end{equation}
In what follows, we work in the canonical ensemble, that is, we consider
variations of the action with constant $\beta $. The variation of the
boundary term is 
\[
\delta B\equiv \delta B_{\phi }+\delta B_{G}\;, 
\]
where 
\begin{equation}
\delta B_{G}=\frac{\beta \sigma }{8\pi G}\left[ N\left( RR^{\prime }\delta
f^{2}-(f^{2})^{\prime }R\delta R\right) +2f^{2}R\left( N\delta R^{\prime
}-N^{\prime }\delta R\right) \right] _{r_{+}}^{\infty }\;,  \label{delG}
\end{equation}
and the contribution from the matter sector is 
\begin{equation}
\delta B_{\phi }=\beta \sigma NR^{2}f^{2}\phi ^{\prime }\delta \phi
|_{r_{+}}^{\infty }\;.  \label{delphi}
\end{equation}
For the black hole solution (\ref{Black-Hole}, \ref{scalar}) the variation
of the fields at infinity are 
\begin{eqnarray}
\left. \delta f^{2}\right| _{\infty } &=&\frac{2G}{l^{2}}\left( G\mu -\frac{%
l^{2}+3G^{2}\mu ^{2}}{r}-\frac{2G\mu l^{2}-8G^{3}\mu ^{3}}{r^{2}}+O\left(
r^{-3}\right) \right) \delta \mu \;, \\
\left. \delta \phi \right| _{\infty } &=&\sqrt{\frac{3G}{4\pi }}\left( \frac{%
1}{r}-\frac{2G\mu }{r^{2}}+O\left( r^{-3}\right) \right) \delta \mu \;, \\
\left. \delta R\right| _{\infty } &=&\left( -\frac{G^{2}\mu }{r}+\frac{%
3G^{3}\mu ^{2}}{r^{2}}+O(r^{-3})\right) \delta \mu \;,
\end{eqnarray}
and thus, we obtain 
\begin{equation}
\left. \delta B_{G}\right| _{\infty }=\frac{3\beta \sigma }{4\pi l^{2}}%
\left( G\mu r-4(G\mu )^{2}-l^{2}/3+O\left( r^{-1}\right) \right) \delta \mu
\;.  \label{B2}
\end{equation}
Note that $\left. \delta B_{G}\right| _{\infty }$ has a potentially
dangerous divergent term. As is shown in Sec. 5, this is a consequence of
the slower fall-off of our metric compared with that of pure gravity with a
standard localized distribution of matter \cite{Henneaux:tv}. This occurs
because the scalar field goes to zero at infinity at a slower rate than
under the usual assumptions. As a result of this there is a non-vanishing $%
\left. \delta B_{\phi }\right| _{\infty }$ given by

\begin{equation}
\left. \delta B_{\phi }\right| _{\infty }=-\frac{3\beta \sigma }{4\pi l^{2}}%
\left( G\mu r-4(G\mu )^{2}+O\left( r^{-1}\right) \right) \delta \mu \;,
\label{B1}
\end{equation}
which exactly cancels the divergence coming from $\left. \delta B_{G}\right|
_{\infty }$ and gives a finite contribution, yielding a finite expression
for the total variation of the boundary term at infinity, 
\begin{equation}
\left. B\right| _{\infty }=-\frac{\beta \sigma }{4\pi }\mu \;.  \label{Binf}
\end{equation}
This is a generic effect observed for scalar fields with a self-interacting
potential unbounded form below in asymptotically AdS\ spacetimes \cite
{Henneaux:2002wm,Henneaux:2004zi,Gegenberg:2003jr,Hertog:2004dr}.

The variation of the boundary term at the horizon, is obtained using 
\begin{eqnarray*}
\left. \delta R\right| _{r_{+}} &=&\delta R(r_{+})-\left. R^{\prime }\right|
_{r_{+}}\delta r_{+}\;, \\
\left. \delta f^{2}\right| _{r_{+}} &=&-\left. (f^{2})^{\prime }\right|
_{r_{+}}\delta r_{+}\;,
\end{eqnarray*}
together with (\ref{regular}), in Eqs. (\ref{delG}, \ref{delphi}). Note that 
$\left. \delta B_{\phi }\right| _{r_{+}}$ vanishes, and hence 
\begin{eqnarray*}
\left. \delta B\right| _{r_{+}} &=&-\frac{\beta \sigma }{16\pi G}%
N(r_{+})\left. (f^{2})^{\prime }\right| _{r_{+}}\delta R^{2}(r_{+}) \\
&=&-\frac{\sigma }{4G}\delta R^{2}(r_{+})\;.
\end{eqnarray*}
Therefore, the boundary term at the horizon is 
\begin{equation}
\left. B\right| _{r_{+}}=-\frac{\sigma }{4G}R^{2}(r_{+})\;.  \label{Bhor}
\end{equation}

Combining Eqs. (\ref{Binf}) and (\ref{Bhor}), the on-shell value of the
Euclidean action is found to be 
\begin{equation}
I=-\frac{\beta \sigma }{4\pi }\mu +\frac{\sigma }{4G}R(r_{+})^{2}\;,
\label{Ieval}
\end{equation}
up to an arbitrary additive constant. Since the Euclidean action is related
to the free energy (in units where $\hbar =k_{B}=1$) as $I=-\beta F$, then 
\begin{equation}
I=S-\beta M\;,  \label{appro}
\end{equation}
where $M$ and $S$ denote mass and entropy, respectively.

Comparing expressions (\ref{Ieval}) and (\ref{appro}), the mass and entropy
are identified as 
\[
M=\frac{\sigma }{4\pi }\mu \;,
\]
and 
\[
S=\frac{\sigma }{4G}R^{2}(r_{+})=\frac{\mbox{Horizon Area}}{4G}\;,
\]
respectively\footnote{%
The mass could also be found using covariant methods, see e.g., \cite{Glenn}.%
}. With these expressions it is simple to check that the first law of
thermodynamics $dM=TdS$ is satisfied.

\section{Phase transition}

The specific heat can be found using Eqs. (\ref{horizon}) and (\ref{beta}) 
\[
C=\frac{\partial M}{\partial T}=\frac{\sigma l^{2}}{4G}\left( \frac{2r_{+}}{l%
}-1\right) \;, 
\]
which is always positive for $r_{+}>l/2$ (\emph{i.e.} $G\mu >-l/4$), and
therefore the black hole dressed with the scalar field can always reach
thermal equilibrium with a heat bath. However, note that for a fixed
temperature, the action principle (\ref{action}), with the same boundary
conditions, also admits the vacuum solution (\ref{phizero}) with $\phi
\equiv 0$. This raises the question of whether one black hole can decay into
the other. Since $Z=\exp (-\beta F)$, this can be examined evaluating the
difference between their respective free energies. The temperature of the
vacuum black hole is 
\begin{equation}
T_{0}=\frac{1}{4\pi }\left( \frac{3\rho _{+}}{l^{2}}-\frac{1}{\rho _{+}}%
\right) \;,  \label{tezero}
\end{equation}
where $\rho _{+}\geq l/\sqrt{3}$. Matching this with the temperature in Eq.(%
\ref{beta}) fixes the relation between their respective horizon radii as 
\begin{equation}
r_{+}=\frac{3\rho _{+}}{4}-\frac{l^{2}}{4\rho _{+}}+\frac{l}{2}\;.
\end{equation}
The Euclidean action evaluated on the vacuum black hole (\ref{phizero})
reads 
\[
I_{0}=\frac{\sigma }{4G}\frac{\rho _{+}^{4}+l^{2}\rho _{+}^{2}}{3\rho
_{+}^{2}-l^{2}}\;, 
\]
and hence, using Eq. (\ref{Ieval}), the difference between both Euclidean
actions is given by 
\begin{equation}
\Delta I=I_{\phi }-I_{0}=\frac{\sigma }{16G}\frac{(l-\rho _{+})^{3}}{\rho
_{+}}\frac{l^{2}+3l\rho _{+}+4\rho _{+}^{2}}{3\rho _{+}^{2}-l^{2}}\;.
\label{DeltaI}
\end{equation}
Analogously, for a fixed temperature, the difference between both black hole
masses is

\begin{equation}
\Delta M=M_{\phi }-M_{0}=-\frac{\sigma }{8\pi G}\frac{(\rho _{+}+l)(\rho
_{+}-l)^{2}}{\rho _{+}^{2}l^{2}}\left[ \rho _{+}^{2}-\frac{l\rho _{+}}{8}-%
\frac{l^{2}}{8}\right] \;,
\end{equation}
which cannot be positive for the allowed range of $\rho _{+}$, $\emph{i.e.}$%
, $\Delta M=M_{\phi }-M_{0}$ $\leq 0$. Similarly, since $S_{0}=\sigma \rho
_{+}^{2}(4G)^{-1}$, the entropies are found to obey $\Delta S=S_{\phi
}-S_{0}\leq 0$. Both inequalities are saturated for $r_{+}=\rho _{+}=l$.
However, at this radius, $\Delta I$ changes of sign, signaling the existence
of a phase transition at the critical temperature 
\[
T_{c}=\frac{1}{2\pi l}\;. 
\]

At the transition temperature, both black hole branches intersect at the
massless configuration (\ref{muzero}) (with $M_{\phi }=M_{0}=0$), describing
a spacetime of negative constant curvature (\emph{i.e.}, locally AdS). The
two phases at each side of the critical point are:

$\mathbf{T>T_{c}}:$

In this phase, $\rho_{+}>l$, and both black holes have positive mass. As $%
\Delta I$ in Eq. (\ref{DeltaI}) is negative, there is a greater probability
for the decay of the black hole dressed with the scalar field into the bare
black hole, induced by thermal fluctuations. In the decay process, the
scalar black hole absorbs energy from the thermal bath, increasing its
horizon radius an consequently its entropy. This suggests that in this
process the scalar field is, in some sense, absorbed by the black hole.

$\mathbf{T<T_{c}:}$

In this phase, $\rho _{+}<l$, both black holes have negative mass, but now $%
\Delta I>0$, which means that the configuration with nonzero scalar field
has greater probability. As a consequence, below the critical temperature,
the bare black hole undergoes a spontaneous ``dressing up'' with the scalar
field. In the process, the mass and entropy of the black hole decrease, so
that the difference in energy and entropy are transferred to the heat bath.

At the critical temperature, the thermodynamic functions of the two phases
match continuously, hence, the phase transition is of second order. The
order parameter that characterizes the transition can be defined in terms of
the value of the scalar field at the horizon, 
\begin{eqnarray*}
\lambda &=&\left| \tanh \sqrt{\frac{4\pi G}{3}}\phi (r_{+})\right| \\
&=&\left\{ 
\begin{array}{lll}
\displaystyle \frac{T_{c}-T}{T_{c}+T} & : & T<T_{c} \\ 
&  &  \\ 
0 & : & T>T_{c}
\end{array}
\right.
\end{eqnarray*}

\section{Summary and discussion}

We have shown the existence of an exact four-dimensional black hole solution
of gravity with a minimally coupled self-interacting scalar field. The
spacetime is asymptotically locally AdS, and has a curvature singularity
hidden by an event horizon which is a surface of negative constant
curvature. The scalar field is regular everywhere outside the origin.

The specific heat is positive for the entire mass range, so that thermal
equilibrium with a heat bath can always be attained. Furthermore, there is a
critical temperature $T_{c}=(2\pi l)^{-1}$. For temperatures above $T_{c}$,
a black hole dressed with the scalar field is likely to decay into the bare
black hole (bare phase). For temperatures below $T_{c}$ however, a bare
black hole would spontaneously acquire a nontrivial scalar field (scalar
phase), in a phenomenon reminiscent of ferromagnetism, where the scalar
field plays the role of the magnetization. It would be interesting to see
how this phase transition manifests itself in the dual CFT \cite
{Witten:1998zw}.

The black hole mass is bounded by $M\ge -\sigma l(16\pi G)^{-1}$, and it is
simple to verify that the strong energy condition is satisfied for $r\geq
r_{+}$, as it occurs for a Schwarzschild-AdS black hole.

The self-interacting potential $V(\phi )$ is negative and unbounded from
below, possessing a global maximum at $\phi =0$. The mass term for scalar
perturbations satisfies the Breitenlohner-Friedman bound that guarantees the
perturbative stability of global AdS spacetime \cite{B-F}. For the topology
considered here, it was shown that the stability of the locally AdS
spacetime (\ref{muzero}) under scalar perturbations, holds provided the mass
satisfies the same BF bound \cite{Aros:2002te}. However, in general, the
stability is no longer guaranteed for non-perturbative solutions. For the
dressed black hole considered here, the scalar field cannot be treated as a
probe. This is because, even for small mass, there is a strong back reaction
that reaches the asymptotic region.

The presence of matter fields with nontrivial asymptotic behavior has
generically two effects: It gives rise to a back reaction that modifies the
asymptotic form of the geometry, and it generates additional contributions
to the charges that depend explicitly on the matter fields at infinity which
are not already present in the gravitational part. These two effects have
been observed in similar setups and for various dimensions in Refs. \cite
{Henneaux:2002wm,Hertog:2004dr,Henneaux:2004zi,Gegenberg:2003jr,BFS}.
Indeed, the scalar field would not contribute to the conserved charges if it
falls off as $\phi \sim r^{-3/2+\varepsilon }$, whereas here the asymptotic
behavior of the scalar field (\ref{scalar}) is 
\[
\phi =\sqrt{\frac{3G}{4\pi }}\left( \frac{\mu }{r}-\frac{G\mu ^{2}}{r^{2}}%
\right) +O\left( r^{-3}\right) .
\]

The effect of this on the geometry can be seen making $r^{2}=\rho
^{2}+G^{2}\mu ^{2}$ in (\ref{Black-Hole}) so that the $g_{\rho \rho }$
component of the metric, which behaves as

\[
g_{\rho \rho }=\frac{l^{2}}{\rho ^{2}}+\left( 1-\frac{3G^{2}\mu ^{2}}{l^{2}}%
\right) \frac{l^{4}}{\rho ^{4}}+O(\rho ^{-5}), 
\]
is the only one relaxed in comparison with the bare black hole, for which

\[
g_{\rho \rho }^{0}=\frac{l^{2}}{\rho ^{2}}+\frac{l^{4}}{\rho ^{4}}+O(\rho
^{-5})\;. 
\]

In spite of the fact that the scalar field is more spread out and the
geometry exhibits a greater deviation from AdS in the asymptotic region, the
conserved charges are still well defined and finite.

\acknowledgments

We thank E. Ay\'on-Beato, G. Barnich, G. Comp\`{e}re and A. Gomberoff for
useful discussions and enlightening comments. C. M. and R. T. wish to thank
Prof. M. Henneaux for his kind hospitality at the Universit\'e Libre de
Bruxelles. This research is partially funded by FONDECYT grants 1010446,
1010449, 1010450, 1020629, 1040921, 7010446, 7010450, and 7020629. The
generous support to Centro de Estudios Cient\'{i}ficos (CECS) by Empresas
CMPC is also acknowledged. CECS is a Millennium Science Institute and is
funded in part by grants from Fundaci\'{o}n Andes and the Tinker Foundation.

\appendix

\section{Appendix}

The form of the self-interacting potential considered here (\ref{potential})
can be naturally obtained through the relation between the conformal and
Einstein frames. Performing a conformal transformation, with a scalar field
redefinition of the form 
\begin{equation}
\hat{g}_{\mu \nu }=\left( 1-\frac{4\pi G}{3}\Psi ^{2}\right) ^{-1}g_{\mu \nu
}\;,\qquad \Psi =\sqrt{\frac{3}{4\pi G}}\tanh \sqrt{\frac{4\pi G}{3}}\phi \;,
\label{ChangingFrames}
\end{equation}
the action (\ref{action}, \ref{potential}) reads 
\begin{equation}
I[\hat{g}_{\mu \nu },\Psi ]=\int d^{4}x\sqrt{-\hat{g}}\left[ \frac{\hat{R}%
+6l^{-2}}{16\pi G}-\frac{1}{2}\hat{g}^{\mu \nu }\partial _{\mu }\Psi
\partial _{\nu }\Psi -\frac{1}{12}\,\hat{R}\,\Psi ^{2}-\frac{2\pi G}{3l^{2}}%
\Psi ^{4}\right] \;.  \label{confaction}
\end{equation}
In this frame, the matter action is invariant under arbitrary local
rescalings $\hat{g}_{\mu \nu }\rightarrow \lambda ^{2}(x)\hat{g}_{\mu \nu }$%
\ and $\Psi \rightarrow \lambda ^{-1}\Psi $, so that the scalar field
equation is conformally invariant.

The black hole solution (\ref{Black-Hole}, \ref{scalar}) acquires a simple
form once expressed in the conformal frame 
\begin{equation}
d\hat{s}^{2}=-\left( \frac{r^{2}}{l^{2}}-\left( 1+\frac{G\mu }{r}\right)
^{2}\right) dt^{2}+\left( \frac{r^{2}}{l^{2}}-\left( 1+\frac{G\mu }{r}%
\right) ^{2}\right) ^{-1}dr^{2}+r^{2}d\sigma ^{2}\;,
\label{Black-Hole-Conformal}
\end{equation}
with 
\begin{equation}
\Psi =\sqrt{\frac{3}{4\pi G}}\frac{G\mu }{r+G\mu }\;.
\label{Scalar-Conformal}
\end{equation}

Note that the mapping between both frames (\ref{ChangingFrames}) is
invertible in the region where the conformal factor $1-4\pi G\Psi ^{2}/3$ is
positive. That is, for nonnegative mass ($\mu \ge 0$), for $r>0$, and for
negative mass, $r>-2G\mu $.

For nonnegative masses this solution possesses only one event horizon and it
has the same causal structure as in the Einstein frame. However, for the
rest of the allowed range, $-l/4<G\mu <0$, the metric develops three
horizons, satisfying $0<r_{--}<-G\mu <r_{-}<l/2<r_{+}$. From a cursory look,
one may say that this solution seems to describe a black hole inside a black
hole. In this frame, the scalar field becomes singular at $r=-G\mu $, but
since the geometry as well as the stress-energy tensor are regular there,
this singularity seems to be harmless. A detailed analysis of the causal and
thermodynamic properties of the solution in the conformal frame is out of
the scope of this appendix, and will be discussed elsewhere.

\end{document}